\documentclass[12pt,preprint]{aastex}
\usepackage{rotating}
\shorttitle{}  
\shortauthors{Katsuda et al.}

\begin{document}

\title{Evidence for Thermal X-Ray Line Emission from the Synchrotron-Dominated Supernova Remnant RX~J1713.7-3946}

\author{Satoru Katsuda\altaffilmark{1}, 
Fabio Acero\altaffilmark{2}, 
Nozomu Tominaga\altaffilmark{3,4}, 
Yasuo Fukui\altaffilmark{5}, 
Junko S. Hiraga\altaffilmark{6}, 
Katsuji Koyama\altaffilmark{7,8}, 
Shiu-Hang Lee\altaffilmark{1}, 
Koji Mori\altaffilmark{9}, 
Shigehiro Nagataki\altaffilmark{10}, 
Yutaka Ohira\altaffilmark{11}, 
Robert Petre\altaffilmark{12},
Hidetoshi Sano\altaffilmark{5}, 
Yoko Takeuchi\altaffilmark{13}, 
Toru Tamagawa\altaffilmark{13}, 
Naomi Tsuji\altaffilmark{14}, 
Hiroshi Tsunemi\altaffilmark{7}, 
\& Yasunobu Uchiyama\altaffilmark{14}
}

\altaffiltext{1}{Institute of Space and Astronautical Science (ISAS), Japan Aerospace Exploration Agency (JAXA), 3-1-1 Yoshinodai, Chuo, Sagamihara, Kanagawa 252-5210, Japan}

\altaffiltext{2}{Laboratoire AIM, CEA-IRFU/CNRS/Universite Paris Diderot, Service d'Astrophysique, CEA Saclay, 91191, Gif-sur-Yvette, France}

\altaffiltext{3}{Department of Physics, Faculty of Science and Engineering, Konan University, 8-9-1 Okamoto, Kobe, Hyogo 658-8501, Japan}

\altaffiltext{4}{Kavli Institute for the Physics and Mathematics of the Universe (WPI), The University of Tokyo, 5-1-5 Kashiwanoha, Kashiwa, Chiba 277-8583, Japan}

\altaffiltext{5}{Department of Physics, Nagoya University, Furo-cho, Chikusa-ku, Nagoya 464-8601, Japan}

\altaffiltext{6}{Department of Physics, School of Science and Technology, Kwansei Gakuin University, Sanda 669-1337, Japan}

\altaffiltext{7}{Department of Earth and Space Science, Osaka University, 1-1 Machikaneyama-cho, Toyonaka, Osaka 560-0043, Japan}

\altaffiltext{8}{Department of Physics, Graduate School of Science, Kyoto University, Kitashirakawa Oiwake-cho, Sakyo-ku, Kyoto 606-8502, Japan}

\altaffiltext{9}{Department of Applied Physics and Electronic Engineering, Faculty of Engineering, University of Miyazaki, 1-1 Gakuen Kibanadai-Nishi, Miyazaki 889-2192, Japan}

\altaffiltext{10}{Astrophysical Big Bang Laboratory, Riken, Wako, Saitama 351-0198, Japan}

\altaffiltext{11}{Department of Physics and Mathematics, Aoyama-Gakuin University, Sagamihara, Kanagawa 252-5258, Japan}

\altaffiltext{12}{Astrophysics Science Division, NASA Goddard Space Flight Center, Greenbelt, MD 20771, USA}

\altaffiltext{13}{High Energy Astrophysics Laboratory, RIKEN Nishina Center, 2-1 Hirosawa, Wako-shi, Saitama 351-0198, Japan}

\altaffiltext{14}{Department of Physics, Rikkyo University, 3-34-1 Nishi-Ikebukuro, Toshima, Tokyo 171-8501, Japan}

%\altaffiltext{13}{National Astronomical Observatory of Japan, Mitaka, Tokyo 181-8588, Japan}
%\altaffiltext{6}{Department of Astronomy, Kyoto University, Kitashirakawa-Oiwake-cho, Sakyo-ku, Kyoto 606-8502, Japan}

%\altaffiltext{4}{Kavli Institute for the Physics and Mathematics of the Universe (WPI), University of Tokyo, 5-1-5 Kashiwanoha, Kashiwa, Chiba 277-8583, Japan}

\begin{abstract}
We report the first detection of thermal X-ray line emission from the supernova remnant (SNR) RX~J1713.7-3946,  the prototype of the small class of synchrotron dominated SNRs.  A softness-ratio map generated using {\it XMM-Newton} data shows that faint interior regions are softer than bright shell regions.  Using {\it Suzaku} and deep {\it XMM-Newton} observations, we have extracted X-ray spectra from the softest area, finding clear line features at $E_{\rm ph}\sim$1\,keV and $\sim$1.35\,keV.  These lines can be best explained as Ne Ly$\alpha$ and Mg He$\alpha$ from a thermal emission component.  Since the abundance ratios of metals to Fe are much higher than solar values in the thermal component, we attribute the thermal emission to reverse-shocked SN ejecta.  The measured Mg/Ne, Si/Ne, and Fe/Ne ratios of 2.0--2.6, 1.5--2.0, and $<$0.05 solar suggest that the progenitor star of RX~J1713.7-3946 was a relatively low-mass star ($\lesssim$20\,M$_\odot$), consistent with a previous inference based on the effect of stellar winds of the progenitor star on the surrounding medium.  Since the mean blastwave speed of $\sim$6000\,km\,s$^{-1}$ (the radius of 9.6\,pc divided by the age of 1600\,yr) is relatively fast compared with other core-collapse SNRs, we propose that RX~J1713.7-3946 is a result of a Type Ib/c supernova whose progenitor was a member of an interacting binary.  While our analysis provides strong evidence for X-ray line emission, our interpretation of its nature as thermal emission from SN ejecta requires further confirmation especially through future precision spectroscopic measurements using {\it ASTRO-H}.  
\end{abstract}

\section{Introduction}

The discovery of synchrotron X-ray emission from the shell of SN~1006 provided the first evidence for diffusive shock acceleration of electrons to very high energies (up to 100 TeV) in SNR shocks, and cemented the long-suspected link between SNRs and cosmic rays \citep{Koyama1995}.  It is now known that almost all young SNRs produce synchrotron radiation in X-rays.  Of these, RX~J1713.7-3946 is the brightest X-ray synchrotron SNR in our Galaxy, making it an exceptionally important target for the study of cosmic-ray acceleration.

RX~J1713.7-3946 is a peculiar SNR in that its X-ray emission is dominated by synchrotron radiation; despite several studies, no thermal line emission has been detected \citep[e.g.,][]{Koyama1997,Slane1999,Uchiyama2003,Cassam-Chenai2004,Takahashi2008,Tanaka2008,Acero2009}.  While \citet{Pannuti2003} and \citet{Hiraga2005} claimed detection of thermal X-ray emission, no emission line features, the most direct evidence for thermal emission, were observed.  Therefore, their claims have been debated.  

The lack of thermal emission can help set an upper limit on the shocked ambient medium density, assuming a plasma geometry and an electron temperature.  Deriving the ambient density is quite important, as it controls the level of the hadronic contribution to the gamma ray emission.  As a matter of fact, the origin of gamma rays (leptonic or hadronic) in several SNRs including RX~J1713.7-3946, has been a matter of considerable debate \citep[e.g.,][]{Aharonian2004,Uchiyama2007,Drury2009,Abdo2011,Ellison2012,Fukui2012,Lee2012,Inoue2012,Gabici2014,Acero2015a}.  \citet{Slane1999} analyzed ASCA data, and estimated the upper limit of the ambient density to be 0.03--0.69\,$D_1^{-0.5}$\,cm$^{-3}$, where $D_1$ is the distance in units of 1\,kpc; the distance of the SNR is well established to be 1\,kpc by the radial velocity of the interacting molecular gas \citep{Fukui2003}, which was subsequently confirmed by \citet{Moriguchi2005} and \citet{Sano2010,Sano2013,Sano2015} from more detailed CO and X-ray studies.  The loose density constraint is mainly due to the unknown plasma temperature, for which \citet{Slane1999} assumed a range of 0.1--2.5\,keV, which would be reasonable given the complicated electron heating mechanism at collisionless shocks \citep[e.g.,][]{Ghavamian2007,Drury2009}.  Later, \citet{Cassam-Chenai2004} analyzed {\it XMM-Newton} data, and placed a much tighter upper limit of 0.02\,$D_1^{-0.5}$\,cm$^{-3}$ at a measured temperature of $\sim$0.5--0.6\,keV.  However, this result is highly uncertain, since the introduction of a thermal component improved the fit only slightly ($\Delta\chi \sim 10-30$).  In fact, the authors did not claim detection of thermal emission. Therefore, non-detection of thermal emission is not sufficient to constrain the ambient density as the upper limit depends strongly on the assumed plasma temperature:  a detection is necessary.

Meanwhile, thermal emission could arise from SN ejecta as well.  In this case, the metal abundances allow us to probe the progenitor star, e.g., its mass and explosive nucleosynthesis within the SN. The ejecta distribution also provides important clues about the explosion mechanism.  In addition, ejecta dynamics and/or the plasma states can reveal the evolutionary state of a SNR.  Thus, the detection of thermal emission from RX~J1713.7-3946 (from either shocked ambient medium or from shocked ejecta) has been sought since its discovery.

Here, we report on the first detection of X-ray line emission from RX~J1713.7-3946, based on our new analysis of  {\it XMM-Newton} and {\it Suzaku} data.  The line emission is detected using both observatories and can be best interpreted as a thermal component from SN ejecta.  Using the inferred abundances and other observational information, we discuss both the progenitor system and the SN type.  

\section{Observations and Data Reduction}

\begin{figure}
\begin{center}
\includegraphics[angle=0,scale=0.8]{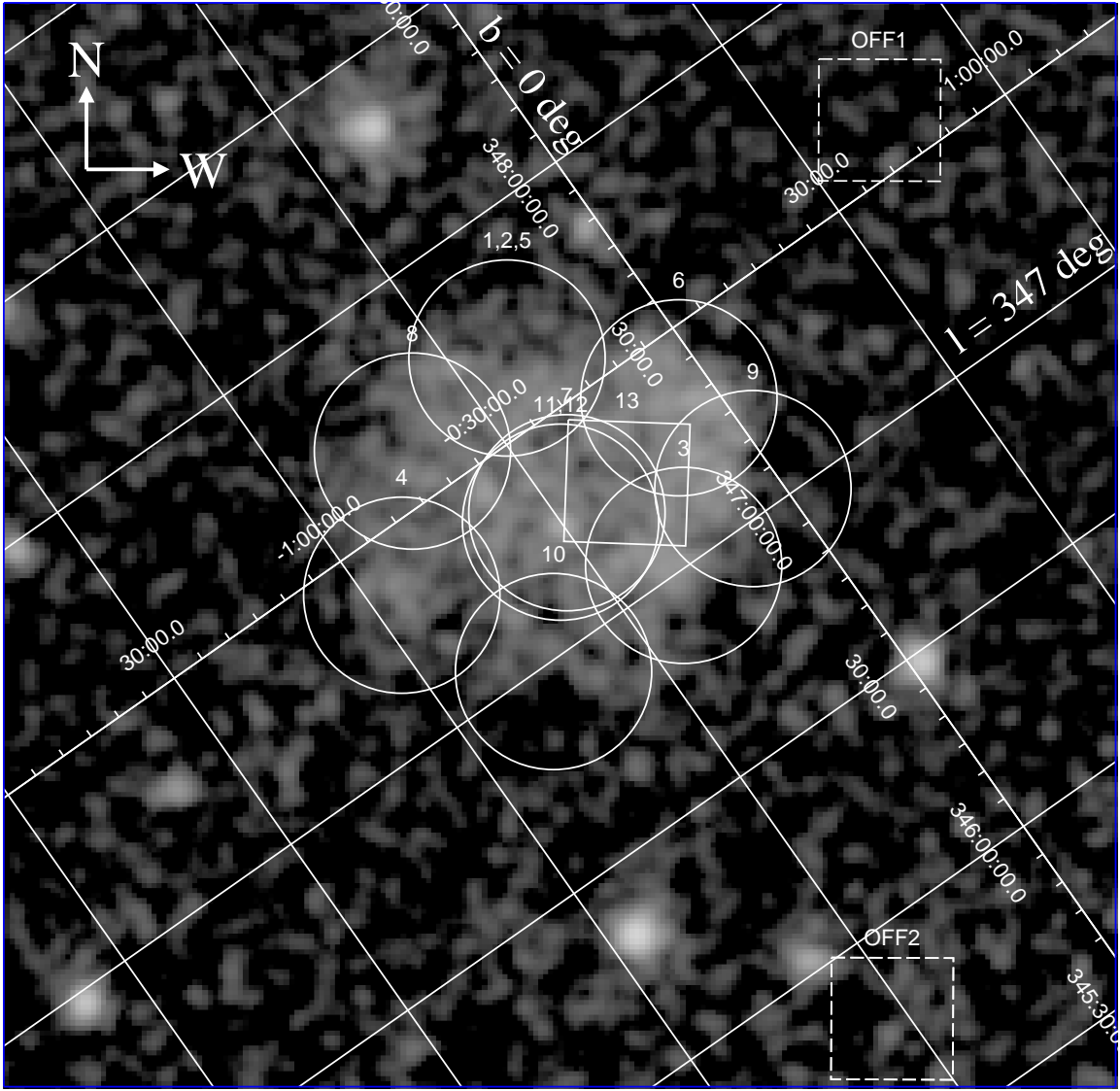}\hspace{1cm}
\caption{The fields of view of the {\it XMM-Newton} (circle) and {\it Suzaku} (box) observations listed in Table~\ref{tab:obs} overlaied on the {\it ROSAT} all-sky survey image.  Position IDs in Table~\ref{tab:obs} are labelled just outside individual fields of view.  The image is binned by 45$^{\prime\prime}$, and has been smoothed with a Gaussian kernel of 1$\sigma$ = 135$^{\prime\prime}$.  
} 
\label{fig:fov}
\end{center}
\end{figure}

We analyzed archival data acquired by {\it XMM-Newton} and {\it Suzaku}, as summarized in Table~\ref{tab:obs}.  The fields of view are shown in Fig.~\ref{fig:fov}.  We reprocessed the raw data, using version 14.0.0 of {\it XMM-Newton} Science Analysis Software and version 20 of the {\it Suzaku} software, together with the latest versions of the calibration files at the analysis phase.  We only used data from the front-illuminated CCDs, since they have slightly better spectral resolution than back-illuminated CCDs:  the MOS camera of {\it XMM-Newton}'s European Photon Imaging Camera \citep[EPIC:][]{Turner2001,Struder2001} and sensors 0, 2, and 3 of {\it Suzaku}'s X-ray Imaging Spectrometer \citep[XIS:][]{Koyama2007}.  Some of the {\it XMM-Newton} data are affected by background (BG) flares.  We thus removed those periods, resulting in the net exposure times given in Table~\ref{tab:obs}.  

To take into account the non X-ray BG (NXB), we used the {\tt xisnxbgen} software \citep{Tawa2008} for the XIS data, while for {\it XMM-Newton} we selected Filter-Wheel-Closed event lists\footnote{${\rm http://xmm2.esac.esa.int/external/xmm\_sw\_cal/background/filter\_closed/mos/index.shtml}$} obtained close in time to our observations.  The total exposure time for the {\it Suzaku} NXB data is ten times that of the exposure time for the observation; for {\it XMM-Newton} the exposure time for the NXB is comparable to that of the observation.  The shorter NXB exposure time for {\it XMM-Newton} is due to the fact that NXB data are relatively scarce.  For our spectral analysis of the central region of the SNR (Section~3), we integrated all available data sets to improve statistics, resulting in merged MOS(1+2) and XIS(0+2+3) spectra with exposure times of 272.2\,ks (plus 259.2\,ks for an eastern part of the central region) and 63.9\,ks, respectively.  The exposure time is at least five times longer than those of previous CCD-based searches for thermal X-rays from the central portion of this remnant.  Each spectrum was grouped into bins with at least 20 counts in order to allow us to use $\chi^2$ statistics.  To fit the spectra, we used the version 12.8.2k of the XSPEC package \citep{Arnaud1996}.

\section{Analysis and Results}

\begin{figure}
\begin{center}
\includegraphics[angle=0,scale=0.7]{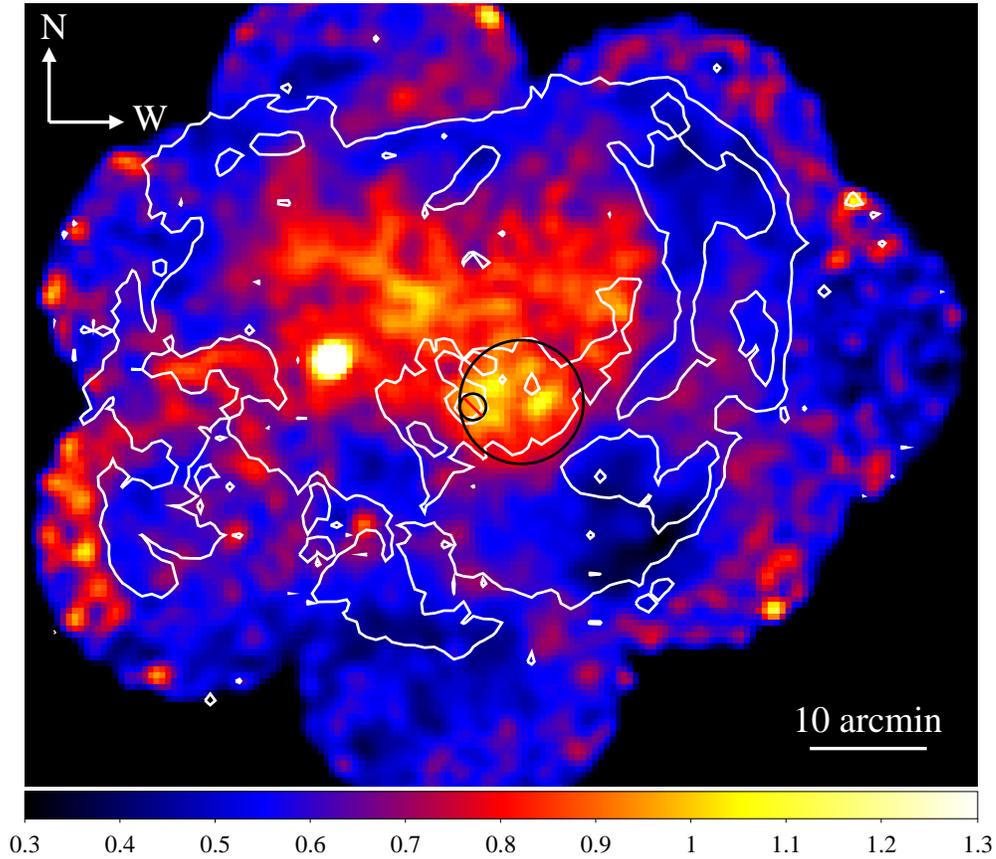}\hspace{1cm}
\caption{Softness (0.5--1.5\,keV/1.5--8\,keV) map generated using {\it XMM-Newton} MOS data.  The image is binned by 30$^{\prime\prime}$, and has been smoothed with a Gaussian kernel of 1$\sigma$ = 90$^{\prime\prime}$.  Overlaid contours represent the X-ray surface brightness.  The black circle is the region where we extract spectra for our spectral analysis.  The CCO was excluded from a black circle with a red line.  The brightest spot seen $\sim$10$^\prime$ east from the CCO is 1WGA~J1714.4-3945, which is believed to be a stellar origin \citep{Pfeffermann1996}.
} 
\label{fig:softness_map}
\end{center}
\end{figure}

\begin{figure}
\begin{center}
\includegraphics[angle=0,scale=0.8]{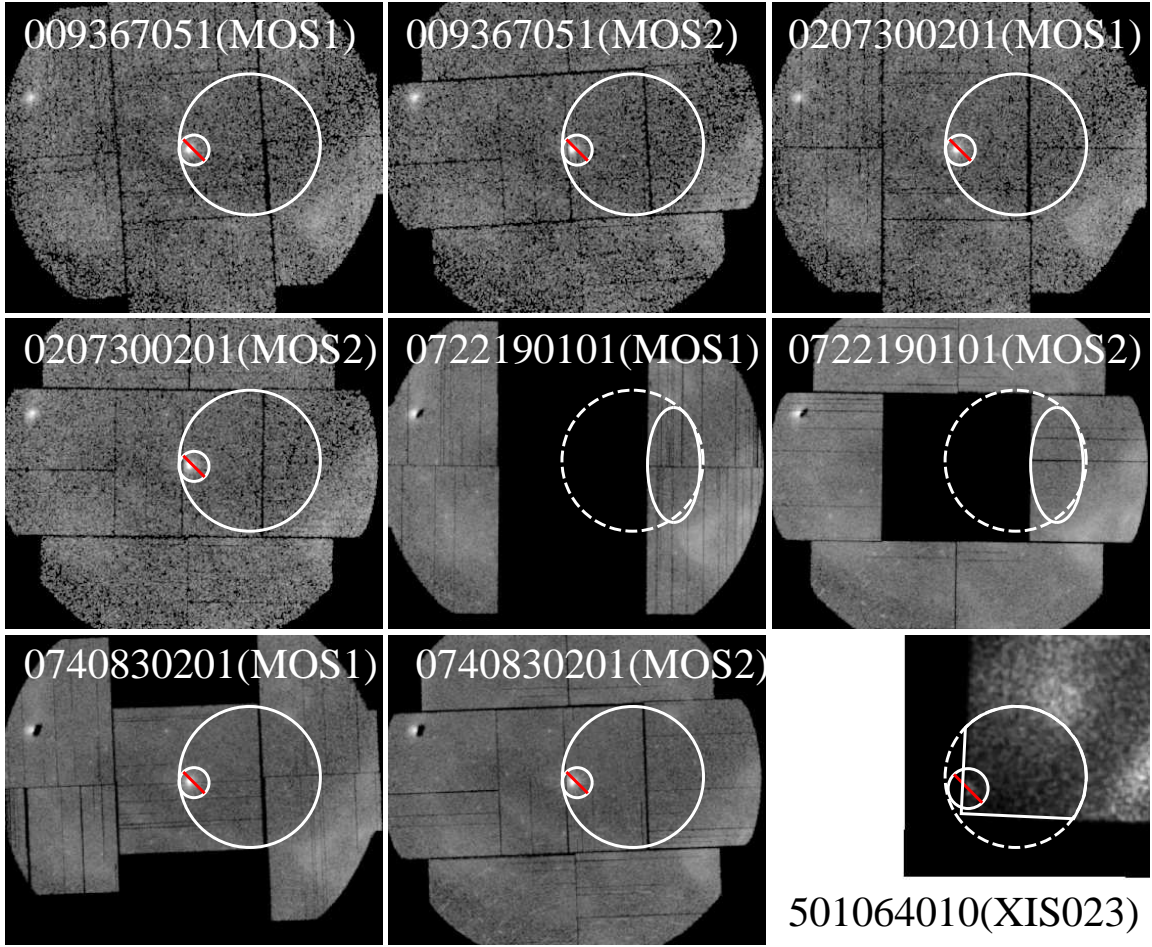}\hspace{1cm}
\caption{Spectral-extraction regions, i.e., solid (partial) circles or ellipses, overlaid on individual images.  The CCO was excluded by the small circles with red lines.  For Obs.IDs 009367051, 0207300201, and 074083021, we merged all the spectra into one long-exposure MOS spectrum, while we did not merge data from Obs.ID 0722190101, since the spectral extraction region covers only western 1/3 area of the circle.  
} 
\label{fig:regs}
\end{center}
\end{figure}

We first generated a softness-ratio map from the {\it XMM-Newton} data, to identify possible locations where thermal X-ray emission might be detectable.  Figure~\ref{fig:softness_map} shows a band-ratio map of 0.5--1.5\,keV to 1.5--8.0\,keV; the low energy band was selected to include the Ne and Mg K-shell lines which are good tracers of thermal emission.  We can see a clear spatial variation:  the interior faint region is softer than the surrounding bright shell region.  This contrast is likely caused by variations of the interstellar absorption, given that the soft region corresponds to a visual extinction hole \citep{Sano2015}.  The soft central region is an ideal location for searching for thermal X-ray emission, because (1) less absorption tends to enhance thermal X-rays which are usually softer than synchrotron X-rays, and (2) the synchrotron X-ray emission is weaker in the interior than in the surrounding shell, making it easier to detect faint thermal line emission in the presence of the dominant synchrotron continuum.  Therefore, we focused our spectral analysis on a circular region enclosing the softest emission, which is shown as a black circle with a radius of 5.3$^{\prime}$ in Fig.~\ref{fig:softness_map}.  The actual spectral extraction regions for the individual {\it XMM-Newton} and {\it Suzaku} observations are shown in Fig.~\ref{fig:regs}.  Note that we excluded the region contaminated by the central compact object (CCO) 1WGA~J1713.4-3946, and that the extraction regions of the Obs.IDs 0722190101 (MOS1+2) and 501064010 (XIS0+2+3) do not exactly match the black circle due to the field of view of those particular observations.  Another (technically) important reason to focus on this area is the existence of a substantial integrated exposure time from the archival {\it XMM-Newton} data, thanks to a recent deep observation of the CCO.  

\begin{figure}
\begin{center}
\includegraphics[angle=0,scale=0.55]{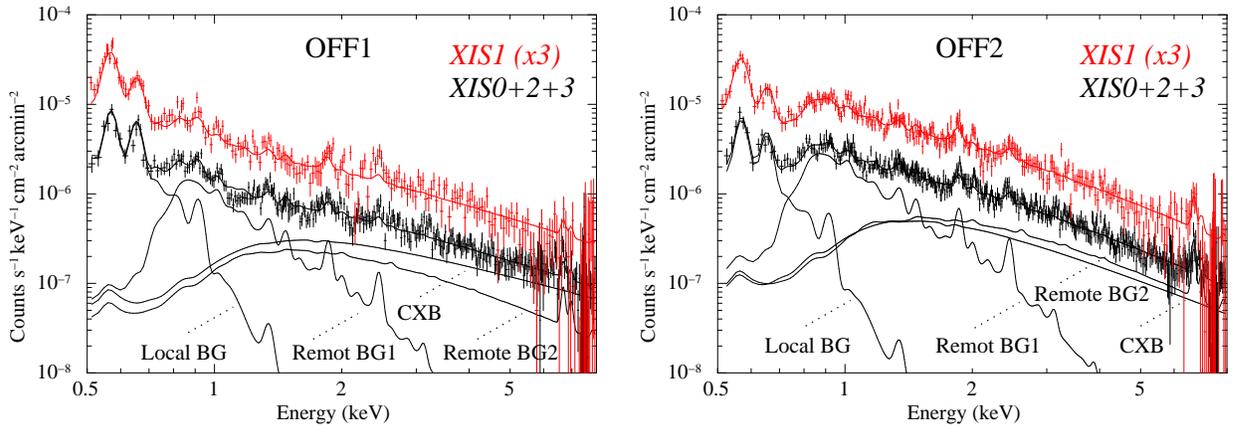}\hspace{1cm}
\caption{Left: NXB-subtracted sky BG spectra from the off-source region 1, together with the individual components of the best-fit model.  Components are shown only for the FI spectrum.  Right: Same as left, but for the off-source region 2.
} 
\label{fig:bg_spec}
\end{center}
\end{figure}

We constructed a sky BG model by fitting NXB-subtracted XIS spectra taken from the two nearby off-source regions, OFF1 and OFF2, which are the only dedicated {\it Suzaku} and {\it XMM-Newton} observations to generate sky BGs of RX~J1713.7-3946 \citep{Takahashi2008,Tanaka2008}.  While there seems to be source-free regions in part of the {\it XMM-Newton} fields of view in Fig.~\ref{fig:fov}, we refrain to use them to avoid possible contaminations of faint diffuse emission associated with RX J1713.7-3946.  Our model consists of an unabsorbed {\tt apec} model to account for the local BG components (local hot bubble and solar-wind charge-exchange emission), and an absorbed {\tt apec} + {\tt apec} + {\tt powerlaw} for the remote BG components (Galactic ridge X-ray emission and cosmic X-ray BG).  This model successfully reproduces both sets as shown in Fig.~\ref{fig:bg_spec}, where individual components are separately shown with labels of Local BG, Remote BG1 (the cooler {\tt apec}), Remote BG2 (the hotter {\tt apec}), and CXB ({\tt powerlaw}).  The best-fit parameters are given in Table~\ref{tab:bg_param}.  We used this model to represent the sky BG.  

\begin{figure}
\begin{center}
\includegraphics[angle=0,scale=0.55]{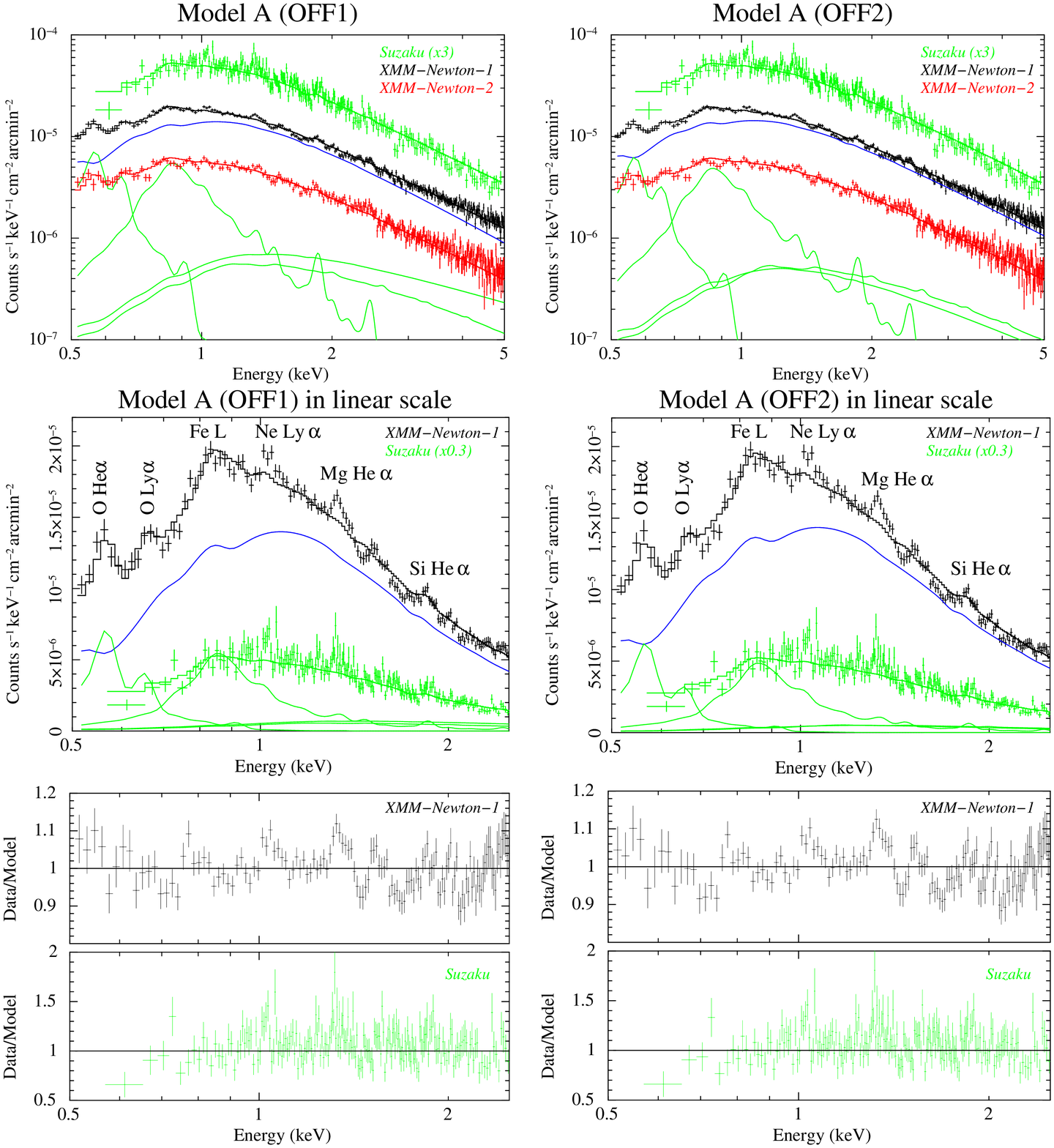}\hspace{1cm}
\caption{Upper panels: NXB-subtracted X-ray spectra from the black circle in Fig.~\ref{fig:softness_map}.  The associated model represents Model A, a power law model (blue) plus sky background (green).  The individual components are shown for the {\it XMM-Newton} data in black.  The left and right columns show the fits using two separate sky background regions OFF1 (left) and OFF2 (right).   
Middle panels: Same as above, but in a linear vertical scale to emphasize the line emission at $\sim$1\,keV, $\sim$1.35\,keV, and $\sim$1.85\,keV.
Lower panels: Residuals associated with the middle panels.  
} 
\label{fig:src_spec1}
\end{center}
\end{figure}

Figure~\ref{fig:src_spec1} shows the NXB-subtracted {\it XMM-Newton} MOS and {\it Suzaku} XIS spectra extracted from the regions in Fig.~\ref{fig:regs}.  The three data sets are a merged, Obs.IDs 0093670501 + 0207300201 + 0740830201, MOS1+2 spectrum in black ({\it XMM-Newton}-1 in the upper panel), a MOS1+2 spectrum from Obs.ID 0722190101 in red ({\it XMM-Newton}-2 in the upper panel), and an XIS0+2+3 spectrum in green.  We simultaneously fit these spectra with a power-law model (Model A) plus the sky BG model whose parameters are listed in Table~\ref{tab:bg_param}.  Since the three spectra are taken from different sized regions as shown in Fig.~\ref{fig:regs}, we allowed their model intensities to vary freely.  For the power-law component, we let the photon index and normalization be free parameters.  For the BG components, we fixed all the parameters at the values in Table~\ref{tab:bg_param}, except for the absorbing column density, which we reduced from $N_{\rm H}\sim1\times10^{22}$cm$^{-2}$ (Table~\ref{tab:spec_param}) to $N_{\rm H}=0.7\times10^{22}$cm$^{-2}$.  This is because our spectral extraction region is located within the absorption hole, where the visual extinction $A_V$ is $\sim$1.5 mag less than that in the surrounding region \citep{Sano2015}, which is equivalent to $\Delta N_{\rm H}$ = 3$\times10^{21}$\,cm$^{-2}$, according to the $N_{\rm H}$--$A_V$ relation, $N_{\rm H}=2.2\times10^{21}$cm$^{-2}$ $A_V$ mag \citep{Gorenstein1975,Watson2011}.  The $N_{\rm H}$ value we employed is consistent with that measured for the CCO, $\sim0.7\times10^{22}$cm$^{-2}$ \citep{Cassam-Chenai2004}.  We allowed the overall normalizations of the local and remote BG components to vary freely, while keeping fixed the relative values of the remote BG components.  

The fit results are shown in Table~\ref{tab:spec_param} (Model A) and Fig.~\ref{fig:src_spec1}.  As is emphasized in the middle and lower panels in Fig.~\ref{fig:src_spec1}, the model does not reproduce the data well, leaving significant residuals around 1\,keV, 1.35\,keV, and 1.85\,keV for the MOS spectra.  Hints of the same residuals can be found in the XIS spectra as well.  Since the residuals appear to be due to the presence of line emission, we refit the data by allowing the abundances of Ne, Mg, and Si in the remote BG low-temperature component to vary.  Allowing these parameters to vary significantly improved the fits.  However, the resulting BG spectral shape is quite different from those in the surrounding off-source regions, and requires unusually high metal abundances.  This result leads us to consider that the thermal emission arises from RX~J1713.7-3946 itself.

\begin{figure}
\begin{center}
\includegraphics[angle=0,scale=0.55]{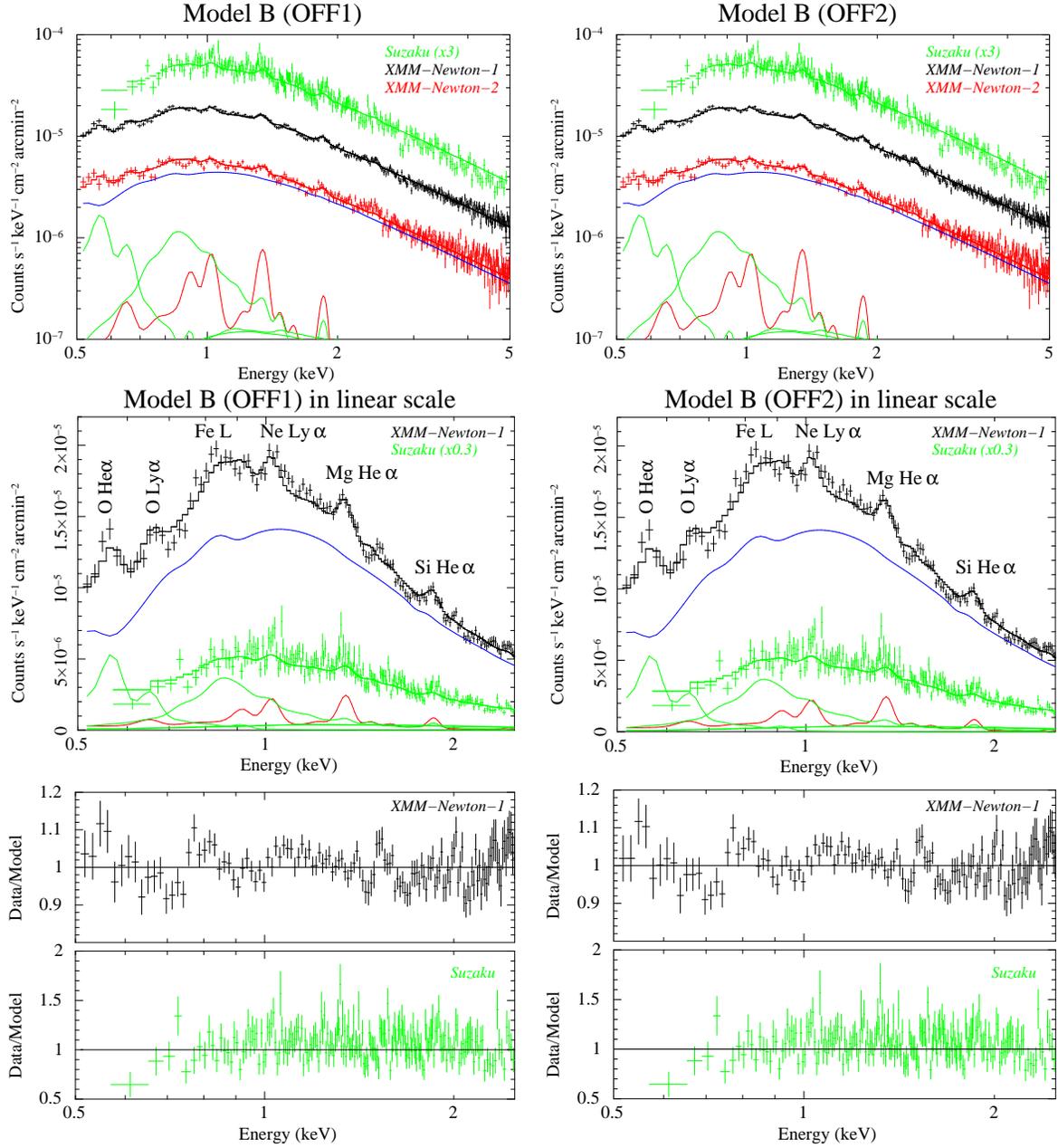}\hspace{1cm}
\caption{Same as Fig.~\ref{fig:src_spec1} but with Model B --- a power law (blue), a thermal component (red), and sky BG (green).
} 
\label{fig:src_spec2}
\end{center}
\end{figure}

To test this possibility, we added a thermal non-equilibrium ionization componetnt ({\tt vvrnei} in the XSPEC) to the power-law model.  This model (Model B) significantly improves the fit, as shown in Fig.~\ref{fig:src_spec2} and Table~\ref{tab:spec_param}, where the fitting details can be found.  The line features at 1\,keV, 1.35\,keV, and 1.85\,keV most likely represent Ne Ly$\alpha$, Mg He$\alpha$, and Si He$\alpha$, respectively.  Note that a bump near 1.5\,keV seen in the {\it XMM-Newton}-1 spectrum would be probably caused by imperfect (over) subtraction of the instrumental BG, specifically the Al-K line at 1.5\,keV.  The statistical significance of adding the thermal component is estimated to be $\gg$99\%, based on the $F$-test; $\chi^2$/d.o.f.\ is reduced from 1312.4/849 to 1148.7/842 when using the OFF1 BG, and from 1330.5/856 to 1154.5/842 for OFF2.  We have verified that the fit results do not change significantly if we use the {\tt vnei} model, which is an older and simpler version of the {\tt vvrnei} model.  

The fact that the line emission is detected not only with the MOS but probably also with the XIS supports that these lines are not related to an instrumental effect.  In addition, the lines do not fall near known instrumental lines such as Al K at 1.5\,keV and Si K at 1.7\,keV in the MOS, ruling out instrumental effects.  On the other hand, the solar wind charge exchange (SWCX) can produce Ne Ly$\alpha$ and Mg He$\alpha$.  \citet{Carter2008} analyzed $\sim$170 {\it XMM-Newton} observations to establish a method for identifying SWCX contamination.  They claimed that SWCX enhancement is characterized by enhanced lines especially from O VII and O VIII.  Since our target (RX~J1713.7-3946) is heavy absorbed by the intervening material, emission below 0.6 keV is dominated by the local BG, i.e., the local hot bubble and/or SWCX.  Therefore, we are able to check whether SWCX enhancement is present in the data using light curves in the 0.5--0.7\,keV energy range that includes O He$\alpha$ and O Ly$\alpha$.  We find no SWCX enhancement in any of the data used in our spectral analysis; all of the light curves for the O K lines are stable except for time periods affected by BG flares due to soft protons.  In addition, the flux below 0.6\,keV (which is dominated by the local BG) is roughly the same between the SNR source spectrum and the BG regions taken outside the SNR (at different epochs), which also suggests no SWCX contamination in our data.

\begin{figure}
\begin{center}
\includegraphics[angle=0,scale=0.55]{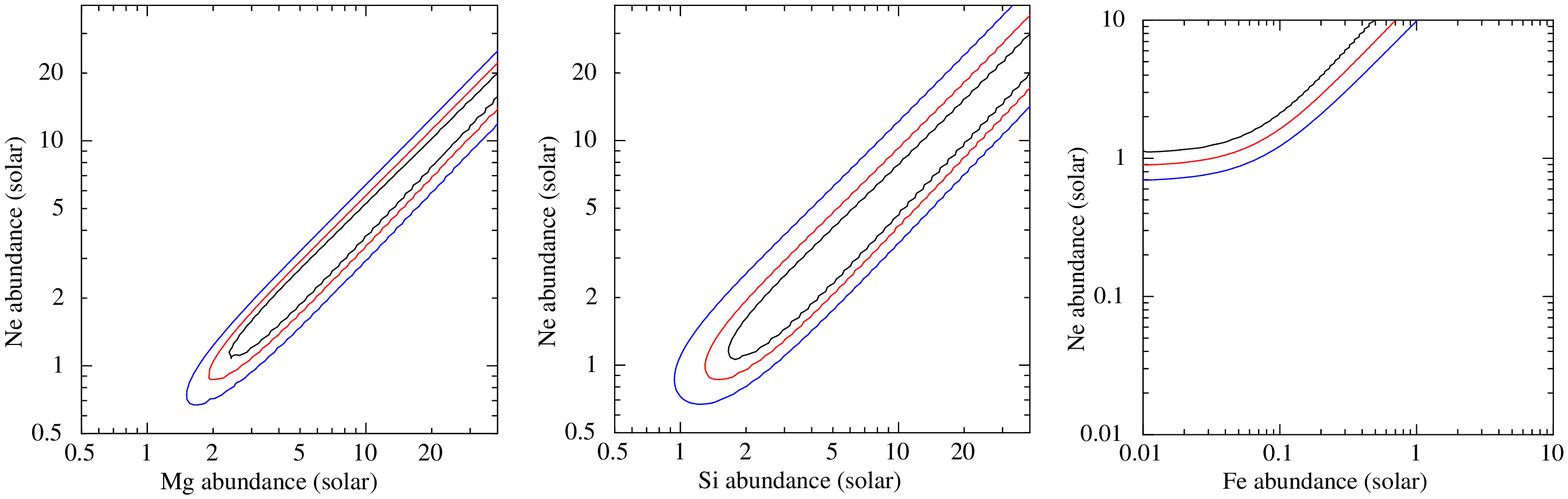}\hspace{1cm}
\caption{Left: Confidence contours for the Ne abundance vs.\ the Mg abundance.  The black, red, and blue contours represent confidence levels of $\Delta \chi^2 = $2.3, 4.6, and 9.2 (corresponding to 68\%, 90\%, and 99\%), respectively.  
Middle: Same as left, but for Ne vs.\ Si.
Left: Same as left, but for Ne vs.\ Fe.
} 
\label{fig:conf_cont}
\end{center}
\end{figure}

The abundance pattern of over-solar Ne, Mg, and Si suggests that the emitting plasma originates from SN ejecta rather than shocked interstellar medium (ISM).  We note that the absolute abundances (X/H) especially for Fe in Table~\ref{tab:spec_param} are poorly constrained.  This is because the thermal continuum from H is hard to measure in an X-ray spectrum dominated by featureless synchrotron radiation, making it difficult to determine absolute abundances.  On the other hand, the relative abundances of metals with prominent emission lines can be constrained more tightly.  Therefore, we created confidence contour plots for Ne vs.\ Mg, Ne vs.\ Si, and Ne vs.\ Fe as shown in Fig.~\ref{fig:conf_cont}.  These are calculated for BG-OFF2, but similar results are found using BG-OFF1.  From these plots, the Mg/Ne ratio is 2.0--2.6 times the solar value with 1-sigma confidence, Si/Ne is 1.5--2.0, and Fe/Ne is $\lesssim$0.05.  These ratios confirm the existence of an Fe abundance deficit, supporting our interpretation that the thermal X-ray emission arises from SN ejecta.  We note that no Fe L line emission associated with this SNR is evident in our spectra; in particular, the 0.7--0.9\,keV band, where prominent 3s$\rightarrow$2p and 3d$\rightarrow$2p lines often dominate SNR spectra, is dominated by both synchrotron radiation and the remote BGs.  This is why we can calculate an upper limit on the Fe/Ne ratio.  Also, the abundance pattern is in agreement with what is expected from a core-collapse (CC) SN explosion, which is evidenced by the presence of the CCO for RX~J1713.7-3946.

We estimate the electron density to be $\sim$0.10--0.13\,cm$^{-3}$ and  the ejecta mass to be $\sim$0.019--0.024\,M$_\odot$, assuming a line-of-sight plasma column depth ($d\ell$) of 3$^{\prime}$ (i.e., 0.87\,pc at a distance of 1 kpc) which corresponds to $\sim$10\% of the shock radius.  Since the area of our spectral extraction region, 84\,arcmin$^2$, corresponds to only $\sim$3\% of the entire remnant, the total ejecta mass in this SNR should be larger than this.  If we simply multiply by the area ratio, then we obtain the total ejecta mass to be $\sim$0.63--0.8\,M$_\odot$.  

\section{Discussion}

\subsection{The Nature of Thermal X-ray Emission}

We have detected X-ray line emission for the first time from the synchrotron-dominated SNR RX~J1713.7-3946, using a long integrated {\it XMM-Newton} exposure as well as a comparatively shorter {\it Suzaku} observation.  This emission can be explained by a thermal component with an emission measure of 2.4--3.9$\times10^{16}$\,cm$^{-5}$ depending on which sky BG we use.  This value is well below the upper limit of $\sim$1$\times10^{17}$\,cm$^{-5}$ at $kT_{\rm e}\sim0.7$\,keV, based on ASCA observations \citep{Slane1999}.  On the other hand, it is inconsistent with estimates from previous detection claims: $\sim$2$\times10^{17}$\,cm$^{-5}$ (at $kT_{\rm }\sim1.6$\,keV) using ROSAT \citep{Pannuti2003} and $\sim$3$\times10^{17}$\,cm$^{-5}$ (at $kT_{\rm }\sim0.6$\,keV) using {\it XMM-Newton} \citep{Hiraga2005}.  These two studies correctly pointed out the presence of thermal X-ray emission from the center of the remnant.  However, in neither instance was line emission detected, so that the analyses suffered from the correlation between the thermal emission and the absorption: the stronger the absorption, the larger the inferred contribution of thermal emission \citep{Hiraga2005}.  This correlation made it difficult to accurately measure the thermal flux.  Therefore, we believe that our result, which is based on a direct detection of thermal line emission, is more reliable than the previous studies.  

Since we have investigated a very limited region (marked as a black circle in Fig.~\ref{fig:softness_map}) of a large SNR, a more thorough search is required to fully characterize the thermal emission.  This is left for future work, presumably based on deep observations using {\it XMM-Newton} and the upcoming {\it ASTRO-H}.  

We measured the abundance ratios of Mg/Ne, Si/Ne, and Fe/Ne of the thermal component to be 2.0--2.6, 1.5--2.0, and $<$0.05, respectively, at the 1-sigma confidence level.  The depleted Fe abundance strongly suggests that the emission originates from SN ejecta rather than swept-up ISM or circumstellar medium (CSM).  We could not detect emission from the swept-up medium.  This is probably because (1) globally, the blastwave is still propagating into a very low density cavity carved by the stellar wind, and (2) in the direction of molecular clouds, the shock-heated plasmas are too cool (due to the high densities) to emit X-rays \citep{Inoue2012,Fukui2012}.  Therefore, in the following discussion, we concentrate on progenitor issues, leaving the physics of particle acceleration to future work, in which we hope to detect thermal emission from the swept-up ISM or CSM.  

\subsection{Inferring a Progenitor Star and the SN Type}

We now use the abundance pattern to constrain the mass of the progenitor star.  The abundance ratios are measured with the X-ray spectra integrated from the contact discontinuity to the reverse shock position along the line of sight, and thus should be compared with nucleosynthetic models in which the yields are integrated from the outside to a certain enclosed mass.  The integrated mass range is chosen to give the best match between the data and each model with different main-sequence masses by chi-square statistics.   Table~\ref{tab:abundance} summarizes abundance ratios of Mg/Ne, Si/Ne, and Fe/Ne expected in CC SN models for main-sequence masses from 13\,M$_\odot$ to 40\,M$_\odot$, for which we employed results from two different groups \citep{Limongi2006,Umeda2005,Tominaga2007,Nomoto2013}.  We examined two cases for each model: (1) the original model derived from the stellar evolutionary calculation of a single star with mass loss, and (2) the CO-core model in which the H and He envelopes are artificially removed, assuming interaction within a close binary system.  From Table~\ref{tab:abundance}, we find that the inferred abundance ratios are more consistent with those of models with low-mass progenitors (M$\lesssim$20\,M$_\odot$) than with high-mass ones (M$\gtrsim$20\,M$_\odot$) for both the original and the CO-core cases.  This is consistent with a previous inference based on a size of the cavity created by the wind from the progenitor star \citep{Cassam-Chenai2004}.  We also note that, for both the H-envelope and CO-core cases, neither the inferred abundance pattern nor the ejecta mass is consistent with what is expected from an electron capture SN, which is a low-energy explosion from a 8--10\,M$_\odot$ star\citep{Wanajo2009}.

\citet{Heger2003} has summarized the results on the endpoint of massive single stars from a theoretical point of view: SNe IIP arise from stars of mass $\sim$9--25\,M$_\odot$ like red supergiants; SNe Ib/c from stars of mass $\gtrsim$35\,M$_\odot$ like Wolf-Rayet stars and other types of core collapse SNe from intermediate mass stars.  Therefore, the indication of low-mass progenitor for RX~J1713.7-3946 suggests a Type IIP origin.  However, a problem occurs in terms of blastwave (or ejecta) velocities.  SNe IIP tend to have slower ejecta velocities than other SNe; early photospheric velocities of most SNe IIP range from $\sim$2000\,km\,s$^{-1}$ to $\sim$6000\,km\,s$^{-1}$ \citep{Spiro2014}, whereas those of stripped-envelope SNe like Type Ib/c usually have blastwave velocities faster than $\sim$10000\,km\,s$^{-1}$ \citep{Wheeler2015}, which can be reasonably understood by considering the presence/absence of a massive H envelope.  For RX~J1713.7-3946, the mean blastwave speed, defined here as the SNR radius over the SNR age, is $\sim$5900\,($\theta$/33$^{\prime}$)\,($d$/1\,kpc)\,($t$/1600\,yr)$^{-1}$\,km\,s$^{-1}$.  The initial blastwave speed must have been faster than this value, since the current shock speed, 3000--4000\,km\,s$^{-1}$ (Tsuji et al.\ 2015 in prep.; Acero et al.\ in prep.), is significantly slower than the mean.  Thus the blastwave speed is more consistent with that of a SN Ib/c than a SN IIP.  Furthermore, we point out that the mean blastwave speed of RX~J1713.7-3946 ($\sim$5900\,km\,s$^{-1}$) is significantly faster than those of similarly-aged CC SNRs: $\sim$2200\,($\theta$/150$^{\prime\prime}$)\,($d$/5\,kpc)\,($t$/1615\,yr)$^{-1}$\,km\,s$^{-1}$ for G11.2-0.3 and $\sim$3200\,($\theta$/22$^{\prime\prime}$)\,($d$/60\,kpc)\,($t$/2000\,yr)$^{-1}$\,km\,s$^{-1}$ for 1E~0102.2-7219 in the SMC (it must be kept in mind however, that this inference depends on the age of RX~J1713.7-3946, which is based on the controversial assumption that it is the result of SN~393 \citep{Fesen2012}).  This discrepancy can be resolved by considering that RX~J1713.7-3946 arose from a Type Ib/c explosion of a relatively low-mass star forming a close binary, in which the (low-mass) progenitor had experienced a binary interaction to remove a massive H envelope.  In fact, both the small Fe/Ne ratio and the small amount of the ejecta mass favor CO-core (i.e., Type Ib/c) models rather than original, single-star ones, supporting the scenario of a Type Ib/c SN from a binary system (although we should be cautious of relatively large systematic uncertainties on these two observables: the Fe abundance is determined from the L-shell lines at 0.7--0.9\,keV whose intensities are very sensitive to the BG level, and the total ejecta mass is inferred from a very small fraction of the remnant, and also we do not know how much SN ejecta have been heated by the reverse shock).

Examples of such a Type Ib/c SN include SN~1994I \citep{Richmond1996} and SN~1993J \citep{Richmond1994}.  Further, the surviving binary companion has been identified in several instances \citep[e.g.,][]{Maund2004}.  There is also growing evidence that a significant fraction of SNe Type Ib/c arise in binary systems; the ratio of Type Ib/c to Type II SNe (0.4$\pm$0.1) with near-solar metallicity is higher than expected from  the initial mass function \citep[e.g.,][]{Nomoto1995}, and no progenitor stars are detected from 10 SNe Ib/c that have deep pre-explosion images, which is difficult to understand if progenitor stars are indeed massive (luminous) stars \citep[e.g.,][]{Smartt2009}. 

Given the possibility of a binary progenitor system, it is interesting to ask if there is a companion star to the CCO in RX~J1713.7-3946.  In a different Galactic SNR, HESS~J1731-347, \citet{Doroshenko2015} recently reported a possible companion star (which used to be associated) with a CCO.  Motivated by this exciting and timely result, we searched for a candidate companion in RX~J1713.7-3946, using archival optical (Palomar Observatory Sky Survey II) and infrared (SPITZER/MIPS) data.  We failed to find an obvious candidate like that in HESS~J1731-347.  However, there are a number of faint stars around the CCO, and further observations are necessary if we are to identify a possible companion star.  By analogy to searches for surviving donor stars in Type Ia SNRs \citep[e.g.,][]{Kerzendorf2014}, key observables would be fast proper motion, large line-of-sight velocity, and abundance anomalies due to contamination by SN ejecta.  Another interesting possibility is that a companion star exploded as a SN earlier than RX~J1713.7-3946, and had already become a neutron star or a black hole when the RX~J1713.7-3946 progenitor exploded.  There is a radio pulsar PSR~J1713-3945 offset from the CCO by $\sim$5 arcminutes \citep{Crawford2002,Pannuti2003}.  Unfortunately, its distance has been estimated to be 4--5\,kpc based on its dispersion measure of 337$\pm$3\,pc\,cm$^{-3}$ \citep{Crawford2002,Cassam-Chenai2004}.  Therefore, the two objects are not likely to be related physically, unless the currently accepted distance to either the CCO or PSR~J1713-3945 (or both) is (are) incorrect \citep[e.g.,][]{Slane1999}.  A black hole scenario remains a possibility.

The guest star in AD393 recorded in Chinese official historical documents has been suggested as the SN that produced RX~J1713.7-3946 \citep{Clark1975,Wang1997}, although this association has been disputed \citep{Fesen2012}.  Its apparent magnitude at maximum brightness would have been around -1 mag, with a plausible upper limit of -3 mag which is set by assuming that the SN was no brighter than Venus (-4 mag).  Combined with the distance of 1\,kpc \citep{Fukui2003}, and the visual extinction of 1.6 at the location of the CCO \citep{Sano2015}, we estimate the absolute magnitude to be -12.6 with an upper limit of -14.6.  This absolute magnitude estimate is consistent with the previous conclusion by \citet{Wang1997}.  One notable fact that \citet{Wang1997} did not explicitly mention is that such a SN is quite faint even for CC SNe, which led \citet{Fesen2012} to claim that SN~393 may not be related to RX~J1713.7-3946.  Our measured low Fe/Ne seems to favor a faint SN scenario, suggesting a faint SN Ib/c for the origin of this remnant.  However, we cannot rule out the possibility that the reverse shock has not yet reached the innermost (Fe-rich) ejecta layer.  In addition, the blastwave speed ($\gtrsim$6000\,km\,s$^{-1}$) inferred above would be unusually fast for faint SNe \citep{Valenti2009}.  On the other hand, we should also be careful about the peak brightness.  In this context, it is still difficult to settle down the discussion on the connection between SN~393 and RX~J1713.7-3946.

\subsection{Future Prospects}

Finally, we comment on the importance of observations with the soon-to-be-launched {\it ASTRO-H} \citep{Takahashi2014}.  {\it ASTRO-H}'s nondispersive Soft X-ray Spectrometer \citep[SXS:][]{Mitsuda2014} will provide additional information that should confirm (or refute) our interpretation, thanks to its unprecedented spectral resolution.  If the thermal emission indeed comes from ejecta, we will find pairs of lines red and blue shifted due to shell expansion, which is not expected for BG emission.  Given the small amount of SN ejecta detected in this SNR, we might be able to find time variation of the abundances, as the reverse shock gradually proceeds back into the unshocked inner ejecta rich in Si and/or Fe.  

We also will use individual line shapes to deduce thermal Doppler broadening, provided information about the acceleration efficiency of cosmic rays at the reverse shock \citep[e.g.,][]{Katsuda2013}.  If efficient cosmic ray acceleration takes place \citep{Drury2009}, we will find narrow lines.  In addition, the accuracy of measurements of abundances, temperature, and ionization timescales will be improved.  

The shell regions are also important locations to be observed with the SXS, since these regions allow us to find thermal emission from the swept-up ISM/CSM.  If thermal emission is detected, we will determine the ambient density, with which we will discriminate between the competing leptonic and hadronic scenarios of the origin of gamma ray emission.  It is also an important task for the SXS to search for thermal line emission with unprecedented sensitivity from other synchrotron-dominated SNRs, e.g., Vela Jr.\ and HESS~J1731-347.

\section{Conclusions}

By analyzing deep {\it XMM-Newton} and relatively shallow {\it Suzaku} observations of the central region of RX~J1713.7-3946, we detected clear X-ray line emission, including Ne Ly$\alpha$ and Mg He$\alpha$, for the first time from this SNR.  Our spectral analysis showed that these lines are likely arising from SN ejecta.  The relative abundances among Ne, Mg, Si, and Fe suggest the progenitor of this remnant was a relatively low-mass star, $\lesssim$20\,M\,$_\odot$.  On the other hand, the blastwave speed is comparatively fast among similarly-aged CC SNRs, which suggests a relatively small amount of SN ejecta, in agreement with what is observed.  This fact, combined with the relatively low-mass progenitor, led us to propose that RX~J1713.7-3946 is the result of SN Ib/c from a close binary system in which binary interactions removed a massive hydrogen envelope.  

\acknowledgments

We thank fruitful discussions with Dr.\ Masaomi Tanaka on supernova types and progenitor stars.  We also thank the anonymous referee for very fast and constructive comments that improved the presentation of the results.  This work is supported by Japan Society for the Promotion of Science KAKENHI Grant Numbers 25800119 (S.Katsuda), 24540229 (K.Koyama), 23000004 (H.Tsunemi).

%\clearpage

\begin{deluxetable}{lccccc}
\tabletypesize{\tiny}
\tablecaption{X-ray data used for our spectral analysis}
\tablewidth{0pt}
\tablehead{
\colhead{Instrument}&\colhead{Obs.~ID (position)}&\colhead{Obs.~Date}&\colhead{Position ($\alpha$, $\delta$)$_{\rm J2000}$}&\colhead{Effective exposure$^a$ (ks)}
}
\startdata
{\it XMM-Newton} MOS1+2$^b$ & 0093670101 (1) & 2001-09-05 & 17:14:11.0, $-$39:25:00 & 17.4 \\
{\it XMM-Newton} MOS1+2$^{b,c}$ & 0093670501 (2) & 2001-09-05 & 17:14:11.0, $-$39:25:44 & 13.5 \\
{\it XMM-Newton} MOS1+2$^b$ & 0093670301 (3) & 2001-09-08 & 17:11:53.0, $-$39:55:59 & 15.6 \\
{\it XMM-Newton} MOS1+2$^b$ & 0093670401 (4) & 2002-03-14 & 17:15:30.0, $-$40:00:57 & 12.1 \\
{\it XMM-Newton} MOS1+2$^{b}$ & 0203470401 (5) & 2004-03-25 & 17:14:11.0, $-$39:25:44 & 16.7 \\
{\it XMM-Newton} MOS1+2$^{b}$ & 0203470501 (6) & 2004-03-25 & 17:11:59.0, $-$39:31:14 & 14.4 \\
{\it XMM-Newton} MOS1+2$^{b,c}$ & 0207300201 (7) & 2004-02-22 & 17:13:23.5, $-$39:48:29 & 16.2 \\
{\it XMM-Newton} MOS1+2$^{b}$ & 0502080101 (8) & 2007-09-15 & 17:15:22.7, $-$39:39:38 & 9.9 \\
{\it XMM-Newton} MOS1+2$^{b}$ & 0502080301 (9) & 2007-10-03 & 17:11:00.9, $-$39:44:26 & 5.4 \\
{\it XMM-Newton} MOS1+2$^{b}$ & 0551030101 (10) & 2008-09-27 & 17:13:32.1, $-$40:11:59 & 24.6 \\
{\it XMM-Newton} MOS1+2$^{c}$ & 0722190101 (11) & 2013-08-24 & 17:13:28.3, $-$39:49:53 & 129.6 \\
{\it XMM-Newton} MOS1+2$^{c}$ & 0740830201 (12) & 2014-03-02 & 17:13:28.3, $-$39:49:53 & 106.4 \\
{\it Suzaku} XIS0+2+3$^b$ & 501064010 (13) & 2006-09-11 & 17:12:39.2, $-$39:43:41 & 21.3 \\
{\it Suzaku} XIS0+1+2+3$^c$ & 100026020 (OFF1) & 2005-09-25 & 17:09:31.9, $-$38:49:24 & 34.9 \\
{\it Suzaku} XIS0+1+2+3 $^c$ & 100026030 (OFF2) & 2005-09-28 & 17:09:05.1, $-$41:02:07 & 37.5 \\
\enddata
\tablecomments{$^a$Averaged among the sensors.  $^b$Used for image analysis. $^c$Used for spectral analysis.}
\label{tab:obs}
\end{deluxetable}

\begin{deluxetable}{lccc}
\tabletypesize{\tiny}
\tablecaption{Spectral-fit parameters of {\it Suzaku}'s off-source regions}
\tablewidth{0pt}
\tablehead{
\colhead{Parameter}&\colhead{OFF1}&\colhead{OFF2}}
\startdata
\multicolumn{3}{l}{Local BG} \\
$kT_{\mathrm e}$ (keV) & 0.19$\pm$0.01 & 0.19$\pm$0.01 \\
$\int n_{\mathrm e} n_{\mathrm H} d\ell$ ($10^{16}$cm$^{-5}$) & 3.54$^{+0.16}_{-0.14}$ & 2.80$\pm$0.13 \\
\hline 
\multicolumn{3}{l}{Remote BG} \\
$N_{\mathrm H}$ (10$^{22}\,$cm$^{-2}$) & 1.0$\pm$0.1 & 1.05$^{+0.04}_{-0.02}$ \\
$kT_{\mathrm e}$1 (keV) & 0.73$\pm$0.03 & 0.74$\pm$0.02 \\
$\int n_{\mathrm e} n_{\mathrm H} d\ell$ ($10^{16}$cm$^{-5}$) & 6.69$^{+0.31}_{-0.38}$ & 15.44$^{+0.38}_{-0.54}$ \\
$kT_{\mathrm e}$2 (keV) & 8.8$^{+3.0}_{-1.7}$ & 7.4$^{+1.0}_{-0.7}$ \\
$\int n_{\mathrm e} n_{\mathrm H} d\ell$ ($10^{16}$cm$^{-5}$) & 4.52$^{+2.88}_{-0.58}$ & 10.53$^{+1.83}_{-0.43}$ \\
$\Gamma$ & 1.25$^{+0.05}_{-0.08}$ & 1.81$^{+0.07}_{-0.06}$ \\
Norm (10$^{-5}$ ph\,keV$^{-1}$\,cm$^{-2}$\,s$^{-1}$\,arcmin$^{-2}$ at 1\,keV) & 1.36$^{+0.07}_{-0.31}$ & 2.95$^{+0.30}_{-0.38}$ \\
\hline
$\chi^{2}$/d.o.f. & 744.3 / 577 & 934.7 / 811 \\
\enddata
\tablecomments{The abundances are set to the solar values.} 
\label{tab:bg_param}
\end{deluxetable}

\begin{deluxetable}{lcccc}
\tabletypesize{\tiny}
\tablecaption{Spectral-fit parameters}
\tablewidth{0pt}
\tablehead{
\colhead{Parameter}&\multicolumn{2}{c}{Model A}&\multicolumn{2}{c}{Model B}}
\startdata
BG region&OFF1&OFF2&OFF1&OFF2 \\
\hline
%\multicolumn{5}{l}{Source emission} \\
$N_{\mathrm H}$ (10$^{22}$\,cm$^{-2}$) & 0.63$^{+0.04}_{-0.02}$ & 0.58$\pm$0.02 
& 0.56$\pm$0.02 & 0.53$^{+0.03}_{-0.01}$\\ 
$\Gamma$ & 2.40$^{+0.06}_{-0.03}$ & 2.27$^{+0.03}_{-0.02}$ 
& 2.23$\pm$0.03 & 2.14$^{+0.02}_{-0.01}$ \\
Norm (10$^{-5}$ ph\,keV$^{-1}$\,cm$^{-2}$\,s$^{-1}$\,arcmin$^{-2}$ at 1\,keV) & 4.23$^{+0.05}_{-0.06}$ & 3.97$^{+0.09}_{-0.06}$ 
& 3.78$^{+0.02}_{-0.07}$ & 3.62$^{+0.03}_{-0.07}$ \\
%{\tt vvrnei} & & & & \\
Initial $kT_{\mathrm e}$ (keV) & --- & --- &   0.1$^{a}$& 0.1$^{a}$ \\
$kT_{\mathrm e}$ (keV) & --- & --- &  0.57$^{+0.27}_{-0.06}$ & 0.58$^{+0.10}_{-0.07}$ \\
log($n_{\mathrm e}t$/cm$^{-3}\,$sec) & --- & --- & 11.61$^{+0.16}_{-0.15}$ & 11.60$^{+0.15}_{-0.18}$ \\
Abundance (solar)\,~~~Ne &  --- & --- & 3.3$^{+4.3}_{-0.7}$ & 2.0$^{+4.0}_{-2.1}$ \\
~~~~~~~~~~~~~~~~~~~~~~~~~~~Mg & --- & --- & 7.0$^{+11.0}_{-4.4}$ & 4.3$^{+4.3}_{-2.1}$ \\
~~~~~~~~~~~~~~~~~~~~~~~~~~~Si & --- & --- & 5.4$^{+47.7}_{-1.6}$ & 3.2$^{+5.1}_{-0.9}$ \\
~~~~~~~~~~~~~~~~~~~~~~~~~~~Fe & --- & --- & $<0.5$ & $<0.4$ \\
$\int n_{\mathrm e} n_{\mathrm H} d\ell$ ($10^{16}$cm$^{-5}$) & --- & --- & 2.39$^{+1.42}_{-0.86}$ & 3.93$^{+7.83}_{-2.28}$ \\
$n_{\mathrm e}$ (cm$^{-3}$) & --- & --- & 0.10 & 0.13 \\
Mass (M$_{\odot}$) & --- & --- & 0.019 & 0.024 \\
\hline 
\multicolumn{5}{l}{Local background} \\
Constant factor & 0.78$^{+0.06}_{-0.07}$ & 0.89$^{+0.11}_{-0.10}$ & 0.66$^{+0.12}_{-0.08}$ & 0.75$^{+0.45}_{-0.07}$ \\
%$\int n_{\mathrm e} n_{\mathrm H} d\ell$ ($10^{16}$cm$^{-5}$) & 2.76$^{+0.22}_{-0.26}$ & 2.51$^{+0.30}_{-0.27}$ & 2.40$^{+0.84}_{-0.20}$ & 2.17$^{+1.14}_{-0.21}$ \\
\hline 
\multicolumn{5}{l}{Remote background} \\
$N_{\mathrm H}$ (10$^{22}$\,cm$^{-2}$) & 0.7$^a$ & 0.7$^a$ & 0.7$^a$&0.7$^a$\\ 
Constant factor & 2.04$^{+0.26}_{-0.21}$ & 0.83$^{+0.07}_{-0.11}$ & 1.40$^{+0.14}_{-0.18}$ & 0.57$^{+0.09}_{-0.16}$ \\
\hline
$\chi^{2}$/d.o.f. & 1312.4 / 849 & 1330.5 / 856  & 1148.7 / 842 & 1154.5 / 842 \\
\enddata
\tablecomments{$^a$Fixed values.  The abundances are set to the solar values unless otherwise stated.  As for BGs, the hidden parameters are the same as those in Table~\ref{tab:bg_param}.}
\label{tab:spec_param}
\end{deluxetable}

% OFF1
% Local APEC: 1.8e-15 erg/s/cm2/arcmin2
% Others: 1.4e-14 erg/s/cm2/arcmin2 vs. 1.5e-14 after setting NH=0.7e22
% OFF2
% Local APEC: 1.3e-15 erg/s/cm2/arcmin2
% Others: 2.0e-14 erg/s/cm2/arcmin2

\begin{deluxetable}{lcccccccccc}
\tabletypesize{\tiny}
\tablecaption{Compositions predicted by CC SN models}
\tablewidth{0pt}
\tablehead{
\colhead{Parameter}&\colhead{Observation}&\colhead{Model}&\colhead{13\,M$_\odot$}&\colhead{15\,M$_\odot$}&\colhead{17/18\,M$_\odot$}&\colhead{20\,M$_\odot$}&\colhead{25\,M$_\odot$}&\colhead{30\,M$_\odot$}&\colhead{40\,M$_\odot$}
}
\startdata
Mg/Ne (solar) & 2.0--2.6 & Umeda --- Original & 1.9 & 2.7 & 2.1 & 0.7 & 0.9 & 0.7 & 0.6 \\
      & & Umeda --- CO core & 2.7 & 5.5 & 2.3 & 0.7 & 1.4 & 0.7 & 1.1 \\
      & & Limongi --- Original & 1.4 & 0.7 & 0.6 & 0.5 & 0.5 & 0.6 & 1.0 \\
      & & Limongi --- CO core & 1.8 & 2.2 & 0.7 & 1.0 & 1.7 & 2.1 & 1.0 \\
\hline
Si/Ne (solar) & 1.5--2.0 & Umeda --- Original & 1.5 & 3.5 & 1.5 & 0.6 & 0.5 & 0.9 & 0.4 \\ 
      & & Umeda --- CO core & 1.6 & 7.7 & 1.3 & 0.5 & 0.2 & 0.8 & 0.1 \\
      & & Limongi --- Original & 1.9 & 1.6 & 1.2 & 0.5 & 0.5 & 0.7 & 0.1 \\
      & & Limongi --- CO core & 1.7 & 1.5 & 0.4 & 0.1 & 0.1 & 0.1 & 0.1 \\
\hline
Fe/Ne (solar) & $<$0.05 & Umeda --- Original & 0.43 & 0.54 & 0.15 & 0.07 & 0.04 & 0.05 & 0.05 \\ 
      & & Umeda --- CO core & 0.006 & 0.005 & 0.009 & 0.018 & 0.047 & 0.026 & 0.010 \\
      & & Limongi --- Original & 0.37 & 0.13 & 0.07 & 0.04 & 0.03 & 0.03 & 0.05 \\
      & & Limongi --- CO core & 0.023 & 0.012 & 0.005 & 0.069 & 0.064 & 0.061 & 0.054 \\
\hline
Mass integrated (M$_\odot$) & $\sim$0.7$^b$ & Umeda --- Original & 10.9 & 12.4 & 15.0 & 16.7 & 19.9 & 22.4 & 19.5 \\
    & & Umeda --- CO core & 0.10 & 0.11 & 1.03 & 1.74 & 0.07 & 4.89 & 0.03 \\
    & & Limongi --- Original & 10.1 & 11.5 & 13.0 & 14.6 & 14.5 & 11.0 & 3.9 \\
    & & Limongi --- CO core & 0.4 & 1.0 & 1.6 & 0.5 & 0.5 & 0.5 & 3.9 \\
\enddata
\tablecomments{$^a$17\,M$_\odot$ and 18\,M$_\odot$ are responsible for Limongi's and Umeda's model, respectively.  $^b$Based on a simple area scaling from the circular region of our interest to the entire remnant.  The abundances are integrated from the outermost envelope to the mass cut, where the residuals between the data and the model minimizes.  The results are based on nucleosynthetic models with solar-metallicity progenitor and explosion energy of $10^{51}$\,erg by \citet{Limongi2006,Umeda2005,Tominaga2007,Nomoto2013}.
}
\label{tab:abundance}
\end{deluxetable}

\end{document}